\documentclass[preprint]{aastex}
\usepackage{longtable}
\usepackage{graphicx}

\slugcomment{Accepted by ApJL}

\shorttitle{ }

\shortauthors{Song et al.}

\begin{document}
\title{Do All Interplanetary Coronal Mass Ejections Have A Magnetic Flux Rope Structure Near 1 AU?}
\author{H. Q. Song\altaffilmark{1}, J. Zhang\altaffilmark{2}, X. Cheng\altaffilmark{3}, G. Li\altaffilmark{4}, Q. Hu\altaffilmark{4}, L. P. Li\altaffilmark{5}, S. J. Chen\altaffilmark{1}, R. S. Zheng\altaffilmark{1}, and Y. Chen\altaffilmark{1}}


\affil{1 Shandong Provincial Key Laboratory of Optical Astronomy
and Solar-Terrestrial Environment, and Institute of Space
Sciences, Shandong University, Weihai, Shandong 264209, China}
\email{hqsong@sdu.edu.cn}

\affil{2 Department of Physics and Astronomy, George Mason
University, Fairfax, VA 22030, USA}

\affil{3 School of Astronomy and Space Science, Nanjing
University, Nanjing, Jiangsu 210093, China}

\affil{4 Department of Space Science and CSPAR, The University of Alabama in Huntsville, Huntsville,
AL 35805, USA}

\affil{5 CAS Key Laboratory of Solar Activity, National Astronomical
Observatories, Chinese Academy of Sciences, Beijing, 100101,
China}





\begin{abstract}
Interplanetary coronal mass ejections (ICMEs) often consist of a shock wave, sheath region, and ejecta region. The ejecta regions are divided into two broad classes: magnetic clouds (MC) that exhibit the characteristics of magnetic flux ropes and non-magnetic clouds (NMC) that do not. As CMEs result from eruption of magnetic flux ropes, it is important to answer why NMCs do not have the flux rope features. One claims that NMCs lose their original flux rope features due to the interactions between ICMEs or ICMEs and other large scale structures during their transit in the heliosphere. The other attributes this phenomenon to the geometric selection effect, i.e., when an ICME has its nose (flank, including leg and non-leg flanks) pass through the observing spacecraft, the MC (NMC) features will be detected along the spacecraft trajectory within the ejecta. In this Letter, we examine which explanation is more reasonable through the geometric properties of ICMEs. If the selection effect leads to different ejecta types, MCs should have narrower sheath region compared to NMCs from the statistical point of view, which is confirmed by our statistics. Besides, we find that NMCs have the similar size in solar cycles 23 and 24, and NMCs are smaller than MCs in cycle 23 but larger than MCs in cycle 24. This suggests that most NMCs have their leg flank pass through the spacecraft. Our geometric analyses support that all ICMEs should have a magnetic flux rope structure near 1 AU.
\end{abstract}



\section{Introduction}
Interplanetary coronal mass ejections (ICMEs) are the counterpart of coronal mass ejections (CMEs) that move out of the solar atmosphere, and can induce strong geomagnetic storms when they interact with the Earth atmosphere (Gosling 1991; Zhang et al. 2007). Studies propose that CMEs result from the eruption of magnetic flux ropes in the solar corona (Chen 2011), which can be formed prior to (Chen 1996; Patsourakos et al. 2013; Cheng et al. 2013) or during (Miki\'{c} \& Linker 1994; Longcope et al. 2007; Qiu et al. 2007; Song et al. 2014; Gopalswamy et al. 2017) an eruption. ICMEs often consist of a shock wave, sheath region, and ejecta region (Kilpua et al. 2013). The ejecta is expected to correspond to the flux rope structure. However, in situ measurements near 1 AU reveal that only about one third ICMEs possess the flux rope features (Chi et al. 2016), which are called magnetic clouds (MC, Burlaga et al. 1981). Most ejecta regions do not have the helical magnetic field feature and can be dubbed as non-magnetic clouds (NMC). Note the MC/NMC ratio has a solar-cycle variation (e.g., Gopalswamy et al. 2008; Chi et al. 2016).

There exist two popular explanations to answer why many ICMEs do not exhibit the flux rope features. One claims that the interactions between ICMEs (Goplaswamy et al. 2001, 2002; Burlaga et al. 2001; Riley et al. 2006) or ICMEs and solar wind (Riley et al. 1997; Odstrcil \& Pizzo 1999) can break up the original flux rope structure of CMEs during their propagation in the heliosphere and make the flux rope feature disappearing, which means the NMC ICMEs do not contain the flux rope near 1 AU. The other one simply attributes the absence of flux rope features to the geometric selection effect (e.g., Zhang et al. 2013), which suggests that the ejecta will exhibit the MC (NMC) features when an ICME has its nose (flank, including leg and non-leg flanks) pass through the observing spacecraft as shown in Figure 1, where the dotted curve represents the shock wave and the solid lines describe the ejecta with the sheath between them.

It is known that the source properties of MCs and NMCs have a good overlap although with differences (e.g., Gopalswamy et al. 2010; Gopalswamy et al. 2013a; Yashiro et al. 2013). However, coronal holes can induce CME deflection that is significant near the Sun (Kay \& Opher 2015). The deflection will move CMEs away from the Sun-Earth line or towards depending on the location of coronal holes relative to the Sun-Earth line and the source regions of CMEs (Gopalswamy et al. 2009, 2012; M{\"a}kel{\"a} et al. 2013). Xie et al. (2013) showed that MCs and NMCs correspond to CMEs that deflected towards and away from the Sun-Earth line, respectively. Beside, the NMCs can also be fitted with flux ropes taking into account of the different impact parameters and adjusting the ejecta boundaries (Marubashi et al. 2015). All of these favor the geometric selection effect when explaining why the NMC features are detected. However, the MC/NMC ratio is lower near the solar maximum (e.g., Chi et al. 2016), which might result from more interactions between ICMEs and support the other explanation.

Now we further address which explanation makes sense from a new point of view that is based on statistics of the geometric size of ICMEs. The CME shocks usually possess a three-dimensional (3D) dome shape in both extreme-ultraviolet and white-light passbands (Veronig et al. 2010; Gopalswamy et al. 2011; Kwon et al. 2014; Liu et al. 2019), and their compression ratios (density ratio of the shock downstream over its upstream) vary along the shock front. In general, the ratios peak around the CME nose and decrease toward the flank portions (Bemporad et al. 2011; Gopalswamy 2011; Gopalswamy et al. 2013b; Kwon et al. 2018). This indicates MCs should have higher shock compression ratio than NMCs. Meanwhile, as the ejecta expansion velocity decreases with time, the separation between the shock and the ejecta first appears in the flank portions, which is observed in both the extreme-ultraviolet (Cheng et al. 2012) and white-light (Song et al. 2019) passbands. Therefore, the widths of sheath region get to the minimum near the nose, which indicates that MCs should have narrower sheath size compared to NMCs. For the ejecta size, the comparison between MCs and NMCs might not be so straightforward as describe below.

Figure 1(a) displays the view observed along the Z axis that is perpendicular to the ecliptic plane (X-Y plane). The red (blue) arrow marks the trajectory of spacecraft passing through the ICME nose (flank). Here the flank is dubbed as leg flank since the spacecraft passes through the ejecta along its leg. In this case, NMCs can have larger or smaller ejecta size compared to MCs, depending on the trajectory within the leg and the cross-section diameter of the ejecta near the nose. Even the spacecraft does not cross the ejecta along its leg, it is still possible to detect both MC and NMC features as shown in Figure 1(b), which presents the view observed along the Y axis that is perpendicular to the X-Z plane. It demonstrates that the impact parameters decide whether the MC or NMC features will be detected. If the trajectory does not pass through the ejecta at all, then only the shock and sheath or just the shock can be detected, also see Figure 2 in Gopalswamy (2006). The red (blue) arrow represents a small (large) impact parameter, indicating the spacecraft trajectory close to the center (periphery) of the flux rope. Here the flank is called non-leg flank as the ejecta leg is not detected by the spacecraft. In this case, NMCs should possess smaller ejecta size than MCs. A statistical comparison between the MC and NMC ejecta sizes can help reveal which kind of flank passing dominates for NMCs.

Besides, if the geometric selection effect holds for all cases, the magnetic field strength of MCs should be stronger compared to NMCs in both sheath and ejecta regions. In this Letter, we conduct the comparative statistics on the geometric and magnetic properties of MC and NMC ICMEs to identify whether the geometric selection works well and reveal which flank passing results in NMCs from the statistical point of view. In Section 2, we introduce the related statistics, and the summary and discussion are given in Section 3.

\section{Statistical results}
\subsection{ICME selection}
To conduct our statistics, ICMEs must be identified first in the background solar wind according to several criteria, such as enhanced magnetic field strength, smoothly changing field direction, low proton temperature and low plasma $\beta$ etc (Zurbuchen \& Richardson 2006; Wu \& Lepping 2011; Song \& Yao 2020). Then each ejecta should be determined to be an MC or NMC. The smooth change of field direction is the most obvious characteristic for MCs. To separate the issues of ICME identification and statistic, and make our study more focused, in this study we select to use the ICME catalog provided by the experts in this field, instead of identifying the events ourselves.

Several complete ICME catalogs are provided based on measurements of the \textit{Advanced Composition Explorer (ACE)} (Richardson \& Cane, 2010, RC catalog), \textit{WIND} (Chi et al. 2016, Chi catalog) and the \textit{Solar Terrestrial Relations Observatory (STEREO)} (Jian et al. 2018, Jian catalog). As no unique and completely objective method can be adopted to judge ICMEs (e.g., Huttunen et al. 2005), some subjective judgements can influence the identifications of ICMEs and their exact boundaries. However, the comparisons between RC and Chi catalogs demonstrate that their results are similar though some differences exist (Chi et al. 2016). This indicates that the discrepancy between different catalogs will not significantly change the statistical results whatever we select the RC or Chi catalog.

The Chi catalog ($http://space.ustc.edu.cn/dreams/wind\_icmes$) not only lists many parameters of each ICME, such as the arrival time of ICME shock, the beginning and ending times of the ejecta, the average velocity of plasma and average magnetic field strength in both the sheath and ejecta regions, but also labels whether the ejecta is an MC or not. At the same time, the catalog also shows each case is an isolated ICME or one of the multiple ICMEs, which is very important for our statistics. As the interaction between multiple ICMEs might greatly influence their structural scale and then field intensity, we avoid the case belonging to the multiple ICMEs. Therefore, the Chi catalog is selected in this study.

\subsection{Comparison between MCs and NMCs}
Totally, 465 ICMEs from 1995 to 2015 are listed in the Chi catalog. After visually examining the catalog, we find 324 isolated ICMEs, and 149 of them have the associated shock, including 76 MCs and 73 NMCs. As Chi catalog does not list the compression ratios of ICME shocks, we resort to a WIND shock catalog analyzed by J. Kasper and M. Stevens ($https://www.cfa.harvard.edu/shocks/wi\_data/$) at the Harvard-Smithsonian Center for Astrophysics, which provides the compression ratios. We get the ratios for 50 MCs and 53 NMCs in total from the shock catalog. Table 1 lists the information for each event, which presents the same contents for both MCs (left) and NMCs (right). Columns (1) and (2) are the arrival date and compression ratio (CR) of each shock, respectively. Columns (3) -- (5) correspond to the sheath size (SS), ejecta size (ES), as well as the size ratio of sheath and ejecta (S/E) sequentially. The average values of total magnetic field strength in the sheath (SBt) and ejecta (EBt) regions are given in columns (6) and (7). Note ``--" indicates no information available.

\begin{center}
\renewcommand\tabcolsep{4.0pt}
\begin{longtable}{ccccccccccccccc}
\caption[Information on the 149 ICMEs (76 MCs and 73 NMCs).]{Information on the 76 MCs and 73 NMCs.}\label{grid_mlmmh} \\
\hline
\hline
\multicolumn{7}{c}{MCs}  & & \multicolumn{7}{c}{NMCs} \\
\cline{1-7}
\cline{9-15}
Date            & CR & SS                 & ES                 & S/E & SBt       & EBt       & & Date            & CR & SS                 & ES                 & S/E & SBt  & EBt \\
\tiny{yy/mm/dd} &    & \tiny{$10^{7}$ km} & \tiny{$10^{7}$ km} &     & \tiny{nT}& \tiny{nT}& & \tiny{yy/mm/dd} &    & \tiny{$10^{7}$ km} &\tiny{$10^{7}$ km}  & & \tiny{nT}&\tiny{nT}\\
\hline
\endfirsthead

\multicolumn{15}{l}%
{{Table 1 (Continued.)}} \\
\hline
\multicolumn{7}{c}{MCs}  & & \multicolumn{7}{c}{NMCs} \\
\cline{1-7}
\cline{9-15}
Date            & CR & SS                 & ES                 & S/E & SBt       & EBt       & & Date            & CR & SS                 & ES                 & S/E & SBt  & EBt \\
\tiny{yy/mm/dd} &    & \tiny{$10^{7}$ km} & \tiny{$10^{7}$ km} &     & \tiny{nT}& \tiny{nT}& & \tiny{yy/mm/dd} &    & \tiny{$10^{7}$ km} &\tiny{$10^{7}$ km}  & & \tiny{nT}&\tiny{nT}\\
\hline
\endhead

\hline \multicolumn{15}{r}{{Continued on next page}} \\ 
\endfoot

\hline \hline
\endlastfoot

  95/03/04 & 1.94 & 1.73 & 2.00 & 0.87 & 06.92 & 11.41 & & 95/03/23 & 1.96 & 1.51 & 2.15 & 0.70 & 08.71 & 08.96\\
  95/08/22 & 2.64 & 1.23 & 2.68 & 0.46 & 06.61 & 09.66 & & 97/05/26 & 1.82 & 0.87 & 2.15 & 0.40 & 08.93 & 10.46\\
  95/10/18 & 2.80 & 1.20 & 4.11 & 0.29 & 09.03 & 21.16 & & 97/09/03 & 1.45 & 0.76 & 1.01 & 0.75 & 13.18 & 13.85\\
  95/12/15 & 2.04 & 1.31 & 3.95 & 0.33 & 06.35 & 10.40 & & 98/02/18 & 1.46 & 2.34 & 3.83 & 0.61 & 14.68 & 08.59\\
  97/01/10 & 2.17 & 0.61 & 3.51 & 0.17 & 08.85 & 14.65 & & 98/06/13 & 3.55 & 0.95 & 2.57 & 0.37 & 09.68 & 09.28\\
  97/02/09 & 2.28 & 2.96 & 2.63 & 1.12 & 06.52 & 07.81 & & 98/08/01 &  --  & 1.10 & 6.24 & 0.18 & 09.02 & 05.71\\
  97/05/15 & 2.17 & 1.32 & 2.23 & 0.59 & 22.02 & 20.45 & & 98/08/10 &  --  & 2.68 & 0.98 & 2.73 & 08.14 & 07.26\\
  97/10/10 & 1.63 & 0.90 & 2.49 & 0.36 & 13.40 & 12.53 & & 98/08/26 & 2.88 & 6.18 & 4.13 & 1.50 & 12.50 & 13.48\\
  97/11/06 &  --  & 2.84 & 3.36 & 0.85 & 13.28 & 15.21 & & 98/10/23 & 2.16 & 1.46 & 4.09 & 0.36 & 09.60 & 06.01\\
  97/11/22 & 2.54 & 1.66 & 3.29 & 0.50 & 23.98 & 16.78 & & 98/11/30 &  --  & 0.81 & 1.33 & 0.61 & 09.75 & 12.62\\
  98/01/06 & 2.83 & 1.81 & 3.94 & 0.46 & 12.13 & 16.13 & & 99/01/22 & 1.60 & 2.12 & 1.92 & 1.10 & 17.55 & 14.60\\
  98/03/04 & 1.43 & 0.50 & 3.79 & 0.13 & 07.41 & 11.08 & & 99/07/02 & 3.18 & 6.64 & 9.14 & 0.73 & 07.85 & 04.43\\
  98/08/19 & 2.30 & 1.69 & 4.02 & 0.42 & 09.98 & 12.92 & & 99/07/08 &  --  & 0.94 & 1.62 & 0.58 & 09.71 & 07.33\\
  98/09/24 & 2.17 & 1.90 & 8.12 & 0.23 & 20.79 & 13.19 & & 99/07/26 & 1.66 & 2.62 & 5.90 & 0.44 & 06.43 & 05.91\\
  98/10/18 &  --  & 1.26 & 3.80 & 0.33 & 12.25 & 17.22 & & 99/09/15 & 1.95 & 0.27 & 1.64 & 0.17 & 12.47 & 12.67\\
  99/04/16 & 2.10 &  --  & 3.75 &  --  & 06.88 & 17.38 & & 99/10/21 & 2.34 & 1.13 & 3.62 & 0.31 & 19.85 & 22.43\\
  00/02/20 & 2.11 & 2.59 & 3.13 & 0.83 & 13.39 & 15.06 & & 99/12/12 & 3.39 & 0.91 & 4.49 & 0.20 & 10.02 & 11.97\\
  00/06/23 & 2.56 & 3.70 & 7.37 & 0.50 & 13.93 & 07.20 & & 00/04/06 & 3.84 & 3.44 & 4.25 & 0.81 & 22.35 & 04.34\\
  00/07/15 &  --  &  --  & 7.33 &  --  & 27.57 & 25.83 & & 00/06/08 &  --  & 2.21 & 7.46 & 0.30 & 17.87 & 10.98\\
  00/07/28 & 2.82 & 0.98 & 3.52 & 0.28 & 18.58 & 14.67 & & 00/07/19 & 3.21 & 3.91 & 3.66 & 1.07 & 10.67 & 06.67\\
  00/09/17 &  --  & 1.82 & 17.9 & 0.10 & 25.80 & 09.65 & & 00/07/26 &  --  & 3.92 & 1.45 & 2.69 & 07.19 & 06.42\\
  00/10/28 & 2.55 & 1.80 & 3.43 & 0.52 & 09.67 & 14.11 & & 00/09/04 & 2.04 & 1.52 & 1.89 & 0.80 & 07.81 & 08.45\\
  01/03/19 & 1.64 & 1.23 & 7.02 & 0.18 & 15.64 & 14.84 & & 00/11/28 & 2.07 & 3.49 & 4.24 & 0.82 & 08.15 & 09.46\\
  01/04/04 & 2.15 & 1.68 & 3.20 & 0.52 & 16.42 & 10.00 & & 00/12/03 & 1.64 & 1.72 & 5.48 & 0.31 & 11.08 & 07.69\\
  01/04/21 & 1.67 & 1.25 & 3.07 & 0.41 & 06.86 & 12.10 & & 00/12/22 & 1.57 & 0.58 & 1.27 & 0.46 & 08.95 & 11.69\\
  01/04/28 & 3.10 & 2.64 & 15.6 & 0.17 & 14.36 & 06.49 & & 01/04/08 & 2.70 & 3.77 & 7.23 & 0.52 & 12.91 & 07.22\\
  01/05/27 & 2.37 & 2.93 & 6.79 & 0.43 & 10.83 & 08.29 & & 01/04/18 & 2.48 & 2.09 & 7.52 & 0.28 & 15.95 & 07.85\\
  01/11/24 & 5.12 & 4.25 & 6.12 & 0.69 & 29.01 & 15.19 & & 01/08/03 & 2.72 & 0.71 & 0.37 & 1.91 & 08.65 & 10.86\\
  02/05/23 & 1.72 & 3.00 & 9.42 & 0.32 & 21.85 & 10.31 & & 01/08/17 & 3.40 & 2.61 & 2.74 & 0.96 & 25.46 & 14.40\\
  02/09/30 & 2.14 & 1.81 & 2.83 & 0.64 & 17.80 & 20.91 & & 01/08/27 & 2.77 & 1.58 & 3.02 & 0.52 & 13.58 & 07.14\\
  03/03/20 &  --  & 2.26 & 2.23 & 1.01 & 10.24 & 11.19 & & 01/10/11 & 2.79 & 2.08 & 0.97 & 2.14 & 19.21 & 20.91\\
  03/11/20 & 4.29 & 0.85 & 3.34 & 0.25 & 23.27 & 30.40 & & 01/12/29 & 3.59 & 2.86 & 2.52 & 1.14 & 16.33 & 16.36\\
  04/04/03 &  --  & 2.93 & 4.65 & 0.63 & 08.66 & 15.59 & & 02/02/28 & 3.89 & 1.90 & 3.84 & 0.50 & 08.68 & 10.15\\
  04/08/29 & 1.88 & 1.47 & 3.83 & 0.38 & 07.17 & 12.06 & & 02/08/18 & 3.52 & 4.69 & 8.40 & 0.56 & 10.34 & 07.70\\
  04/11/07 & 2.27 & 2.51 & 6.91 & 0.36 & 43.71 & 10.53 & & 02/09/19 &  --  & 3.51 & 4.45 & 0.79 & 06.64 & 04.72\\
  04/11/09 & 3.30 & 0.61 & 5.35 & 0.11 & 23.82 & 25.01 & & 02/10/02 & 2.08 & 1.19 & 5.41 & 0.22 & 09.74 & 11.11\\
  05/05/15 & 4.92 & 0.94 & 20.8 & 0.05 & 17.02 & 14.25 & & 02/11/16 &  --  & 1.71 & 5.60 & 0.31 & 09.60 & 09.43\\
  05/05/20 &  --  & 0.56 & 3.70 & 0.15 & 09.87 & 12.06 & & 03/05/09 &  --  & 1.47 & 9.14 & 0.16 & 10.60 & 07.34\\
  05/06/12 &  --  & 1.25 & 3.61 & 0.35 & 21.19 & 14.28 & & 04/04/26 &  --  & 1.94 & 2.49 & 0.78 & 09.79 & 05.81\\
  05/06/14 & 2.13 & 2.16 & 4.68 & 0.46 & 09.35 & 08.96 & & 04/07/26 & 3.61 & 3.06 & 4.28 & 0.72 & 18.29 & 22.81\\
  05/07/17 & 1.61 & 2.46 & 0.98 & 2.52 & 09.06 & 13.14 & & 04/08/01 &  --  & 2.47 & 2.57 & 0.96 & 06.92 & 05.48\\
  06/04/13 &  --  & 0.81 & 3.61 & 0.23 & 10.25 & 16.91 & & 04/09/13 &  --  & 4.31 & 4.37 & 0.98 & 11.87 & 06.72\\
  06/12/14 & 3.53 & 3.04 & 3.58 & 0.85 & 13.19 & 13.25 & & 04/11/11 &  --  & 3.61 & 4.67 & 0.77 & 09.89 & 07.16\\
  07/11/19 & 1.93 & 0.96 & 2.20 & 0.44 & 07.17 & 17.47 & & 05/05/07 & 1.48 & 4.73 & 12.2 & 0.39 & 11.73 & 07.14\\
  09/02/03 &  --  & 0.63 & 2.12 & 0.30 & 09.24 & 10.09 & & 05/07/10 & 1.74 & 1.33 & 6.36 & 0.21 & 22.89 & 11.65\\
  09/06/27 & 1.55 & 1.11 & 3.00 & 0.37 & 05.49 & 07.57 & & 05/08/24 & 2.50 & 2.84 & 4.65 & 0.61 & 27.50 & 08.46\\
  10/04/05 &  --  & 1.41 & 5.53 & 0.26 & 15.22 & 08.78 & & 05/09/02 & 2.23 & 1.43 & 2.01 & 0.71 & 14.05 & 09.01\\
  10/05/28 &  --  & 2.37 & 2.70 & 0.88 & 06.53 & 13.48 & & 05/09/15 & 2.76 & 1.95 & 6.45 & 0.30 & 10.43 & 05.88\\
  10/08/03 & 2.35 & 3.50 & 2.91 & 1.20 & 12.46 & 08.86 & & 06/07/09 & 2.33 & 3.34 & 2.83 & 1.18 & 06.11 & 09.24\\
  11/02/18 & 4.19 & 3.67 & 6.27 & 0.58 & 15.03 & 09.64 & & 06/08/19 & 1.80 & 4.05 & 3.83 & 1.06 & 13.56 & 08.04\\
  11/03/29 &  --  & 1.24 & 3.20 & 0.39 & 06.94 & 12.86 & & 06/12/16 & 1.90 & 1.84 & 3.78 & 0.49 & 09.23 & 03.69\\
  11/05/28 &  --  & 0.91 & 3.04 & 0.30 & 10.70 & 11.47 & & 10/02/10 & 1.49 & 1.51 & 1.39 & 1.09 & 07.93 & 06.97\\
  11/06/04 & 2.75 & 1.02 & 3.02 & 0.34 & 20.12 & 10.94 & & 10/04/11 & 1.96 & 1.98 & 1.72 & 1.15 & 08.70 & 11.19\\
  11/10/05 & 1.61 & 0.50 & 1.45 & 0.35 & 09.00 & 11.74 & & 11/02/04 &  --  & 1.05 & 1.67 & 0.63 & 05.43 & 13.06\\
  11/10/24 &  --  & 1.26 & 2.06 & 0.61 & 17.00 & 22.34 & & 11/08/05 & 2.26 & 5.75 & 5.11 & 1.13 & 13.07 & 04.49\\
  12/02/26 &  --  & 3.55 & 3.22 & 1.10 & 06.81 & 13.07 & & 11/11/28 & 2.09 & 0.87 & 0.95 & 0.91 & 12.27 & 15.19\\
  12/06/16 & 1.72 & 0.47 & 2.30 & 0.20 & 23.40 & 28.81 & & 12/01/22 & 2.13 & 2.77 & 5.81 & 0.48 & 19.17 & 07.25\\
  12/10/08 & 1.87 & 1.84 & 2.83 & 0.65 & 12.83 & 15.58 & & 12/03/08 &  --  & 4.04 & 10.6 & 0.38 & 14.67 & 07.14\\
  12/10/12 &  --  & 1.91 & 2.61 & 0.73 & 05.57 & 10.44 & & 12/03/15 &  --  & 2.27 & 3.33 & 0.68 & 12.47 & 08.02\\
  12/10/31 & 2.33 & 1.14 & 3.35 & 0.34 & 09.65 & 11.35 & & 12/04/23 & 2.24 & 2.00 & 1.36 & 1.47 & 11.20 & 14.48\\
  12/11/12 & 2.04 & 1.46 & 2.55 & 0.57 & 14.62 & 21.17 & & 12/07/14 & 2.08 & 2.83 & 5.76 & 0.49 & 12.82 & 19.53\\
  12/11/23 & 2.12 & 2.24 & 2.95 & 0.76 & 11.27 & 11.73 & & 12/11/26 & 2.09 & 3.12 & 6.47 & 0.48 & 07.26 & 04.19\\
  13/04/13 & 2.34 & 3.29 & 3.82 & 0.86 & 20.24 & 10.21 & & 13/03/17 & 2.87 & 4.35 & 7.31 & 0.59 & 11.90 & 08.76\\
  13/05/25 &  --  & 0.75 & 5.48 & 0.14 & 10.29 & 06.84 & & 13/08/22 &  --  & 5.35 & 4.53 & 1.18 & 04.57 & 06.84\\
  13/06/27 & 2.42 & 2.11 & 3.40 & 0.62 & 07.73 & 11.22 & & 14/02/15 & 2.12 & 2.35 & 1.43 & 1.64 & 11.52 & 14.72\\
  13/07/12 & 1.98 & 2.17 & 6.20 & 0.35 & 08.94 & 12.30 & & 14/04/20 & 3.30 & 4.89 & 4.82 & 1.02 & 08.63 & 05.61\\
  13/10/02 & 3.13 & 4.82 & 2.66 & 1.81 & 09.93 & 07.97 & & 14/06/07 & 2.09 & 4.76 & 7.03 & 0.68 & 12.83 & 04.00\\
  14/08/19 &  --  & 1.57 & 4.63 & 0.34 & 10.26 & 16.52 & & 15/06/16 &  --  & 2.15 & 3.23 & 0.67 & 06.82 & 06.12\\
  14/08/26 &  --  & 2.26 & 2.32 & 0.98 & 05.13 & 12.54 & & 15/08/07 &  --  & 1.66 & 4.20 & 0.39 & 08.05 & 09.06\\
  14/09/12 &  --  & 1.45 & 8.14 & 0.18 & 21.32 & 17.65 & & 15/08/15 & 2.31 & 2.18 & 1.83 & 1.20 & 14.94 & 11.36\\
  15/01/02 &  --  & 1.27 & 0.86 & 1.48 & 07.21 & 09.03 & & 15/09/07 &  --  & 2.72 & 5.56 & 0.49 & 10.00 & 15.73\\
  15/01/07 &  --  & 0.35 & 2.01 & 0.18 & 15.23 & 18.24 & & 15/10/24 & 2.90 & 3.15 & 9.85 & 0.32 & 07.70 & 05.76\\
  15/03/31 &  --  & 1.50 & 2.29 & 0.66 & 13.64 & 12.35 & & 15/12/31 & 2.56 & 2.83 & 2.89 & 0.98 & 12.22 & 12.97\\
  15/05/06 & 2.18 & 2.61 & 3.39 & 0.77 & 13.95 & 12.71 & &          &      &      &      &      &       &     \\
  15/05/18 &  --  & 0.48 & 0.55 & 0.86 & 15.29 & 18.10 & &          &      &      &      &      &       &     \\
  15/11/06 &  --  & 2.87 & 6.37 & 0.45 & 15.40 & 15.85 & &          &      &      &      &      &       &     \\
\end{longtable}
\end{center}

Figure 2(a1) presents the probability density functions (PDF) of the shock compression ratios for MCs (red) and NMCs (blue). The ratios are sorted into 0.3 bins and both PDFs are very similar. However, the similarity may change with the choice of bin size. For this reason, it is preferable to use the cumulative distribution function (CDF) or survival function (SF, i.e., 1-CDF), in which the bin size is irrelevant. The SF equates to the fraction of shocks whose compression ratios exceed a given value and is more intuitive to interpret. For example, 32\% of MCs have a compression ratio over 2.5, whereas it is 37\% for NMCs as shown in Figure 2(a2). Therefore, the SF is adopted in our study.

A Kolmogorov-Smirnov (KS) test is used to verify whether the two compression ratio distributions of MCs and NMCs are from the same probability distribution p. The KS test is a non parametric and distribution free test, which makes no assumption about the distribution of data. Its null hypothesis is that the two distributions do indeed come from p, and the alternative hypothesis is that they do not. A p-value higher than 0.05 is not statistically significant and indicates weak evidence against the null hypothesis. The p-value 0.803 means the compression ratios of MCs and NMCs follow the identical distribution. Further more, they still have the same average value 2.43. In a word, we do not find any significant differences between the distributions of MC and NMC shock compression ratios. This seems to conflict with the expectation of geometric selection effect and will be discussed in Section 3. Note that the numbers and corresponding average values of MCs (red) and NMCs (blue), and the p-value (green/black when beyond/below 0.05) are written in each SF panel of Figures 2 -- 4.

Figures 2(b1) and (b2) display the same analyses with the shock compression ratios but for the sheath sizes. Figure 2(b1) shows the PDFs for 74 MCs (2 events with data gap are removed) and 73 NMCs with a bin of 4 $\times$ $10^{6}$ km, and the SFs are displayed in Figure 2(b2). They demonstrate that the difference between distributions of MC and NMC sheath sizes is significant. The p-value of the KS test is 0.005 (less than 0.05), confirming that the sheath sizes of MCs and NMCs follow different distributions. The average value of NMC sheath sizes is 2.54 $\times$ $10^{7}$ km, larger than MCs (1.80 $\times$ $10^{7}$ km) by $\sim$41\%. This is consistent with the previous expectation of geometric selection effect.

Figures 2(c1) and (c2) illustrate the analytical results of the ejecta sizes for 76 MCs and 73 NMCs with a bin of 1 $\times$ $10^{7}$ km in the PDF panel. The average values of MC (4.36 $\times$ $10^{7}$ km) and NMC (4.18 $\times$ $10^{7}$ km) ejecta sizes are close, and the p-value is 0.074. Therefore, no obvious differences exist between the ejecta size distributions of MCs and NMCs. This seems to support that most NMCs result from the leg flank passing of the spacecraft, and will be further analyzed later.

Figures 2(d1) and (d2) show the PDFs and SFs of sheath/ejecta ratios for both MCs and NMCs. As not all ICMEs have the same size, this ratio is a normalization to the spatial scales of ICMEs, and can remove the bias of size. The bin size for this ratio is 0.15. The difference between MC and NMC ratios is significant with a p-value being 0.004. The average values are 0.54 and 0.77 for MCs and NMCs, respectively. As MCs have narrower sheath region, it makes sense that they have smaller sheath/ejecta ratios. This is also consistent with the geometric selection effect.

The narrower sheath region of MCs agrees with the expectation of the geometric selection effect as compression is expected to peak at the ICME nose, which also predicts that the magnetic field strength in the sheath region of MCs should be stronger than that of NMCs. Meanwhile, as the magnetic field strength in the central part of flux ropes is larger than that in the outer portion, MCs should also have stronger ejecta field compared to NMCs. Figure 3 presents the sheath and ejecta field strength distributions of two types of ejecta with a bin of 3 nT. Panels (a1) and (b1) illustrate the analytical results of the sheath field for 76 MCs and 73 NMCs. The average value of MCs (13.53 nT) is larger than that of NMCs (11.73 nT) by $\sim$15\%, while their distributions are similar with a p-value being 0.054. For the ejecta field intensity, the average value of MCs (13.81 nT) exceeds that of NMCs (9.59 nT) by $\sim$44\%, and their distributions are totally different (p$=$0) as shown in Panels (a2) and (b2). This confirms that MCs and NMCs have different magnetic properties, consistent with previous studies (e.g., Gopalswamy et al. 2018).

\subsection{Comparison between MCs and NMCs individually for two solar cycles}
Previous statistics found that properties of MCs vary in different solar cycles. Compared to cycle 24, MCs in cycle 23 possess stronger magnetic field intensities in both the sheath and ejecta regions, and they also exhibit larger spatial size (Gopalswamy et al. 2015). We compare the spatial size and magnetic field strength of MCs and NMCs individually for two cycles to check whether some variations exist compared to Subsection 2.2. Table 1 shows that 40 MCs (50 NMCs) belong to cycle 23 (1996-2008), and 32 MCs (22 NMCs) cycle 24 (2009-2015).

Figures 4(a1) and (a2) present the SFs of sheath size of MCs and NMCs in cycles 23 and 24, respectively, which demonstrate that NMCs possess thicker sheath region than MCs in both cycles, consistent with the result in Figure 2. Figures 4(b1) and (b2) show that MCs in cycle 23 have larger size (5.28$\times$$10^{7}$ km) than MCs in cycle 24 (3.35$\times$$10^{7}$ km), consistent with Gopalswamy et al. (2015), and the MCs are larger than NMCs in cycle 23, consistent with Gopalswamy et al. (2018). However, the ejecta sizes of NMCs in both cycles are relatively similar, and MCs have smaller size (3.35$\times$$10^{7}$ km) than NMCs (4.40$\times$$10^{7}$ km) in cycle 24 as shown in Figure 4(b2).

As displayed in Figure 1, whatever the ICMEs have their leg or non-leg flank pass through the spacecraft, the NMCs could exhibit smaller size than MCs. However, only passing through the leg flank can make NMCs have larger ejecta size than MCs. Therefore, the NMCs in cycle 24 should mainly result from spacecraft passing through ICMEs along the ejecta leg. Although the MC size is larger in cycle 23 compared to cycle 24, each ejecta has the same leg length near 1 AU. NMCs in both cycles have the similar ejecta sizes, which implies that the NMCs in cycle 23 might also mainly result from spacecraft passing through the leg flank of ICMEs, similar to cycle 24 intrinsically. Thus we suggest that most NMCs should correspond to the leg portion of flux ropes. Figures 4(c1)-(d2) exhibit the magnetic field intensities in both sheath and ejecta regions. For both cycles, MCs have stronger fields than NMCs in both regions, while MCs and NMCs have the same (different) distributions of sheath (ejecta) fields, consistent with Figure 3.

\section{Summary and discussion}
In this Letter, we conducted some comparative statistics on several parameters, including the shock compression ratio, the sheath and ejecta sizes, the sheath/ejecta ratio, as well as the magnetic field strength in both sheath and ejecta regions between 76 MCs and 73 NMCs from 1995 to 2015. We also compared the spatial sizes and magnetic field intensities of MCs and NMCs individually for two solar cycles. Our main results include: (1) MCs have narrower sheath region and stronger magnetic field than NMCs, agreeing with the expectation of geometric selection effect and supporting that all ICMEs have the flux rope structure near 1 AU. (2) NMCs in both cycles have the similar ejecta size, which is smaller than MCs in cycle 23 but larger than MCs in cycle 24. Our analyses suggest that NMCs mainly result from the spacecraft passing through the ICMEs from the leg flank.

As mentioned, MCs do not exhibit larger shock compression ratio than NMCs as expected by the geometric selection effect. The individual comparisons of compression ratios for two cycles present the similar results (not shown). We give a simple discussion about it. Though the ICME velocity near the nose is faster, while the background density and magnetic field intensity should also influence the compression ratio. The shock covers a large spatial scale near 1 AU, and the compression ratio does not decrease with strict monotonicity from the nose to the flank but with some fluctuations (see Figure 9 of Kwon et al. 2018). Therefore, the detected compression ratio near the nose could be beyond or below the ratio near the flank as the measurement is conducted only through a single satellite for each ICME, and the compression ratio can not be employed to verify the geometric selection effect for limited events.

\acknowledgments We thank the referee for the comments and suggestions that influenced our original manuscript significantly. We acknowledge the use of ICME catalog provided by Dr. Yutian Chi et al. at the University of Science and Technology of China, and the shock catalog from Dr. Michael L. Stevens and Prof. Justin C. Kasper at the Harvard-Smithsonian Center for Astrophysics. Hongqiang Song is grateful to Drs. Pengfei Chen (NJU), Ying Liu (NSSC), Chenglong Shen (USTC), Fang Shen (NSSC), and Liang Guo (a statistician at SDU) for their helpful discussions. This work is supported by the Shandong Provincial Natural Science Foundation (JQ201710), the NSFC grants U1731102, U1731101, and 11790303 (11790300), as well as the CAS grants XDA-17040507.


\clearpage

\begin{figure}
\epsscale{0.9} \plotone{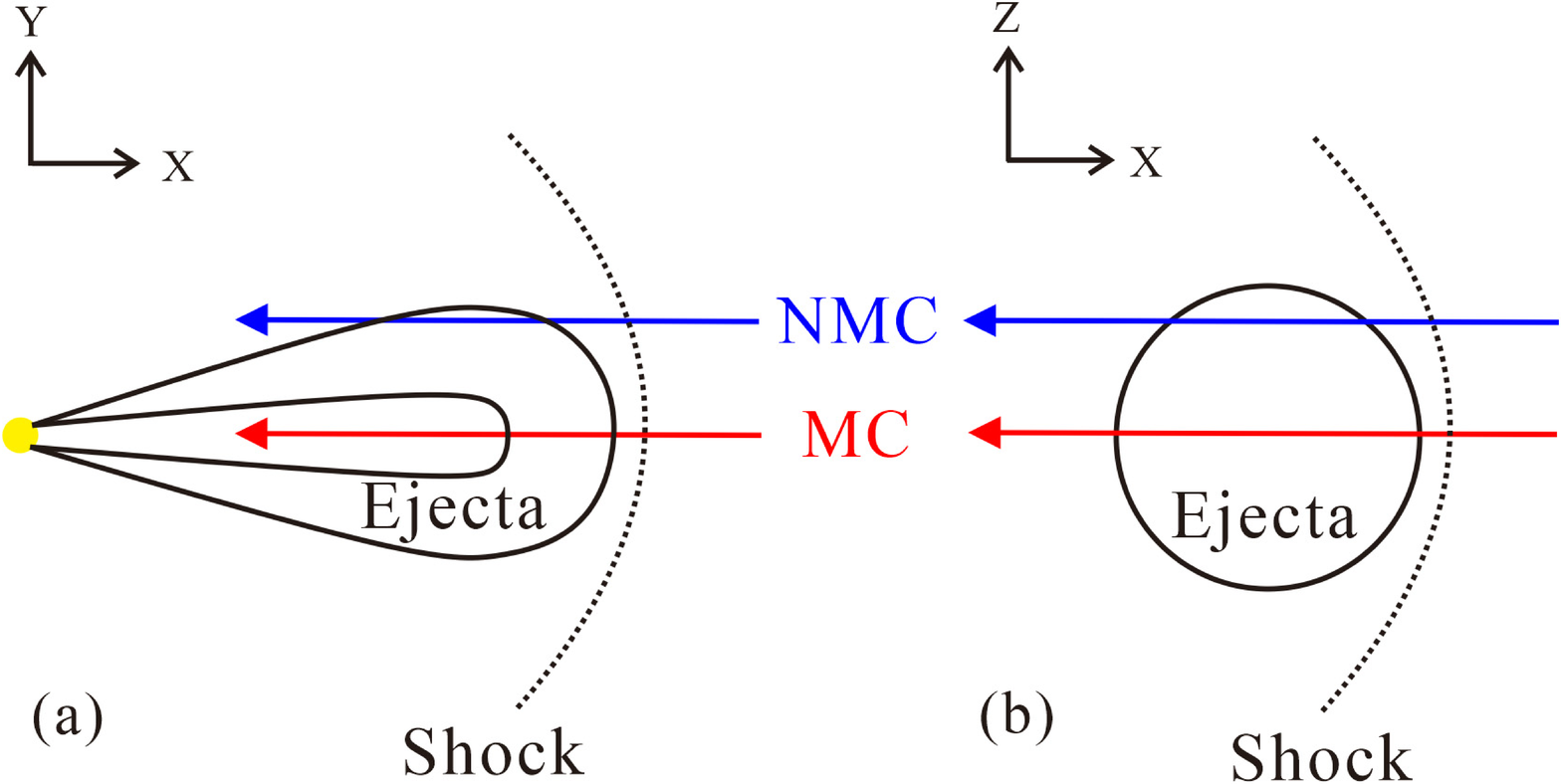} \caption{Schematic drawing of the geometric selection effect. The dotted curves and the solid lines represent the shock wave and ejecta, respectively. The red/blue arrow denotes the spacecraft trajectory throughout the ICME, along which the MC/NMC features will be detected. \label{Figure 1}}
\end{figure}

\begin{figure}
\epsscale{0.9} \plotone{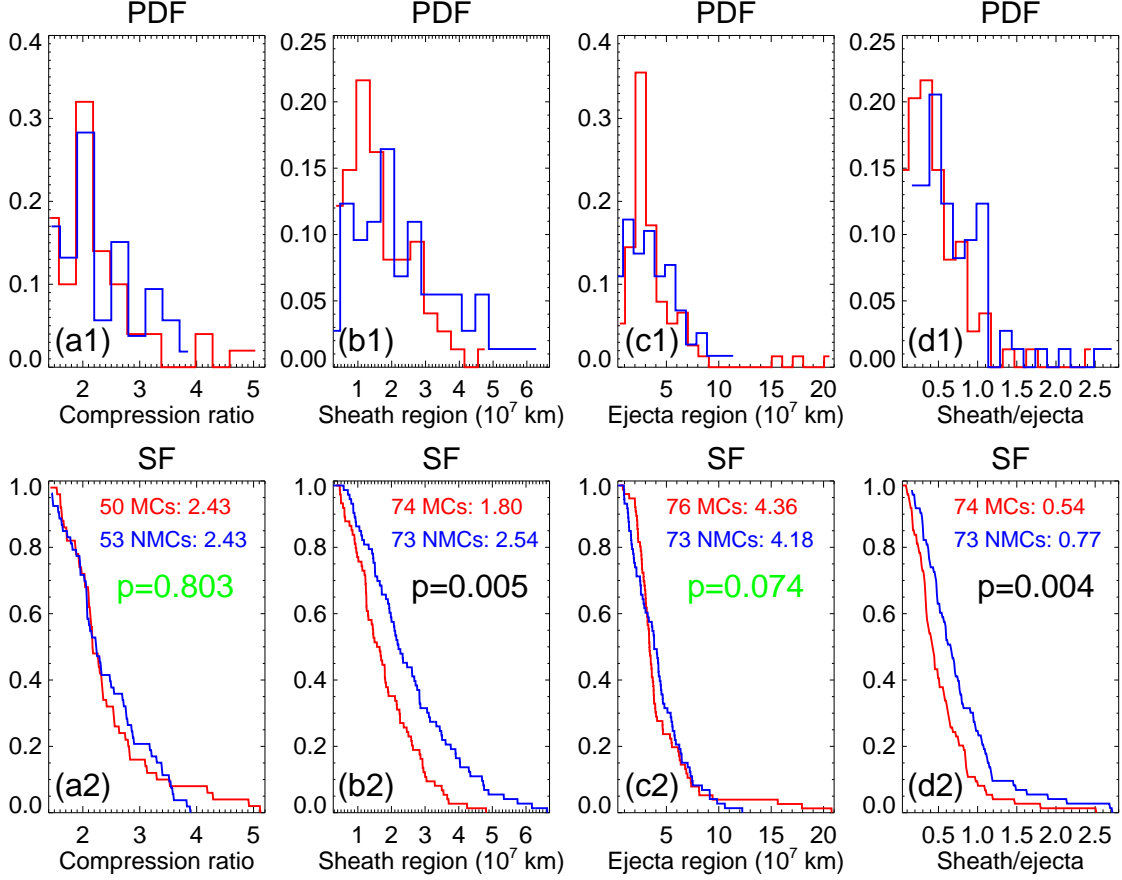} \caption{The PDFs (top) and SFs (bottom) of shock compression ratio (a), sheath size (b), ejecta size (c), as well as sheath/ejecta ratio (d) for MCs (red) and NMCs (blue). The PDFs use bins of 0.3, 4$\times$$10^{6}$ km, 1$\times$$10^{7}$ km, and 0.15 sequentially. KS test comparing the MC and NMC distributions shows that the difference are marginal for the compression ratio and ejecta size. The difference is significant in the case of the sheath size and sheath/ejecta ratio. \label{Figure 2}}
\end{figure}

\begin{figure}
\epsscale{0.5} \plotone{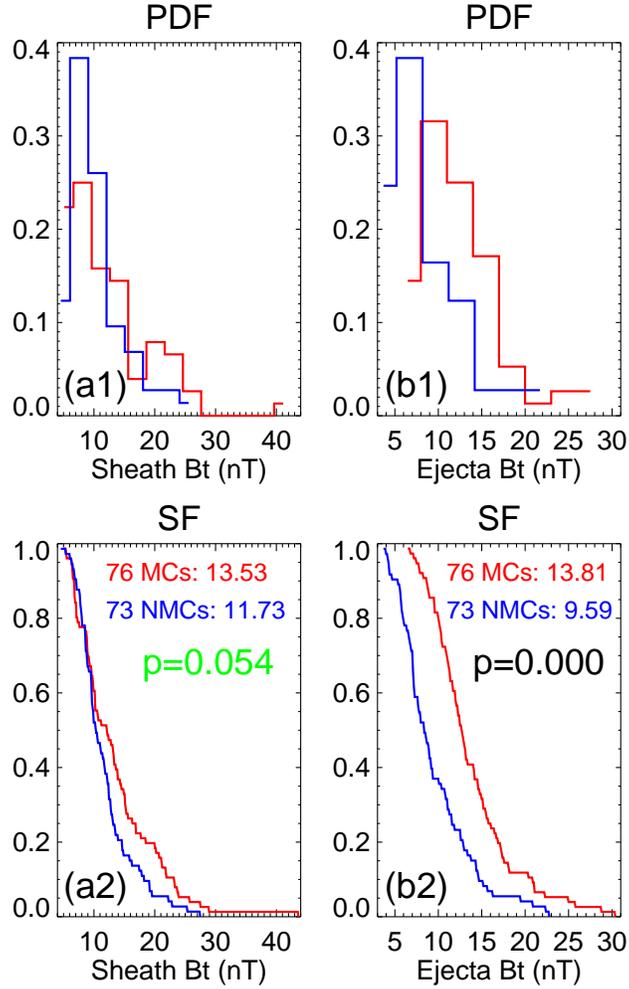} \caption{The PDFs (top) and SFs (bottom) of magnetic field intensities in sheath (a) and ejecta (b) for MCs (red) and NMCs (blue). The PDF uses a bin of 3 nT. The KS test shows that the difference is marginal for the sheath field and significant for the ejecta field. \label{Figure 3}}
\end{figure}

\begin{figure}
\epsscale{0.9} \plotone{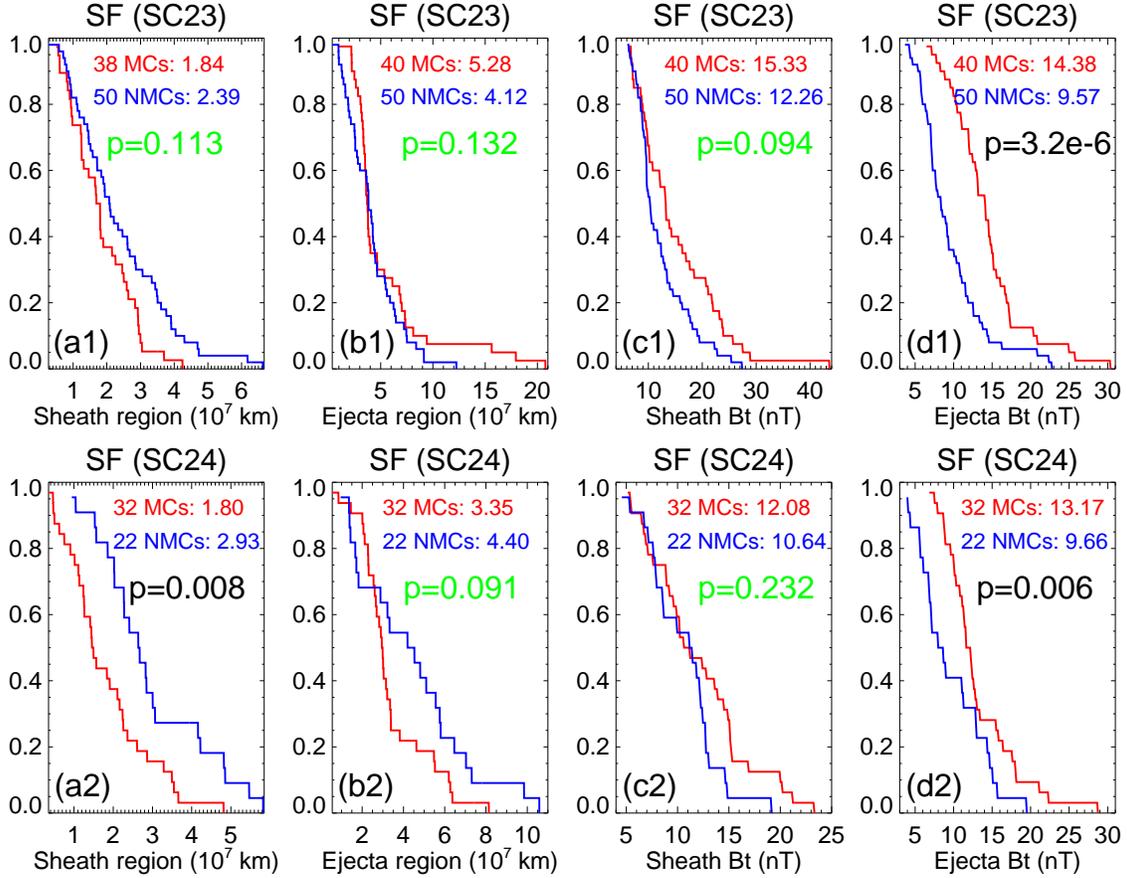} \caption{The SFs of sheath size (a), ejecta size (b), as well as magnetic field intensities of sheath (c) and ejecta (d) in solar cycle 23 (top) and 24 (bottom) for MCs (red) and NMCs (blue). The KS test shows that the difference is significant for the sheath size in cycle 24 and magnetic field intensity in both cycles. \label{Figure 4}}
\end{figure}





\begin{thebibliography}{}






\bibitem[Bemporad et al. (2011)]{Bemporad 2011 shock ratio}
Bemporad, A., \& Mancuso, S. 2011, \apjl, 739, L64




\bibitem[Burlaga et al. (1981)]{Burlaga 1981}
Burlaga, L., Sittler, E., Mariani, F., \& Schwenn, R. 1981, \jgr,
86, 6673

\bibitem[Burlaga et al. (2001)]{Burlaga 2001}
Burlaga, L., Skoug, R., Smith, C., et al. 2001, \jgr, 106, 20957







\bibitem[Chen (1996)]{Chen 1996}
Chen, J. 1996, \jgr, 101, 27499



\bibitem[Chen(2011)]{2011LRSP....8....1C}
Chen, P.~F.\ 2011, LRSP, 8, 1












\bibitem[Cheng et al.(2013)]{2013ApJ...769L..25C}
Cheng, X., Zhang, J., Ding, M.~D., et al.\ 2013, \apjl, 769, L25


\bibitem[Cheng et al. (2012)]{Cheng 2012 wave separation from the ejecta}
Cheng, X., Zhang, J., Olmedo, O., et al. 2012, \apj, 745, L5


\bibitem[Chi et al. (2016)]{ChiYutian 2016 ICME catalog}
Chi, Y. T., Shen, C. L., Wang, Y. M., et al. 2016, \solphys, 291, 2419


























\bibitem[Gopalswamy (2006)]{Gopalswamy 2006}
Gopalswamy, N. 2006, \ssr, 124, 145

\bibitem[Gopalswamy (2011)]{Gopalswamy 2011}
Gopalswamy, N. 2011, in Proc. of the 7th International Workshop on Planetary, Solar and Heliospheric Radio Emissions, ed. H. O. Rucker et al. (Graz: \"Osterreichische Akademie der Wissenschaften Verlag), 325

\bibitem[Gopalswamy et al. (2010)]{Gopalswamy 2010}
Gopalswamy, N., Akiyama, S., Yashiro, S., \& M{\"a}kel{\"a}, P. 2010, in Magnetic Coupling between the Interior and the Atmosphere of the Sun, ed. S. S. Hasan \& R. J. Rutten (Berlin: Springer), 289

\bibitem[Gopalswamy et al. (2008)]{Gopalswamy 2008}
Gopalswamy, N., Akiyama, S., Yashiro, S., Michalek, G., \& Lepping, R. P. 2008, JASTP, 70, 245

\bibitem[Gopalswamy et al. (2018)]{Gopalswamy 2018}
Gopalswamy, N., Akiyama, S., Yashiro, S., \& Xie, H. 2018, JASTP, 180, 35

\bibitem[Gopalswamy et al. (2012)]{Gopalswamy 2012}
Gopalswamy, N., M{\"a}kel{\"a}, P., Akiyama, S., et al. 2012, \jgr, 117, A08106

\bibitem[Gopalswamy et al. (2013)]{Gopalswamy 2013a}
Gopalswamy, N., M{\"a}kel{\"a}, P., Akiyama, S., et al. 2013a, \solphys, 284, 17

\bibitem[Gopalswamy et al. (2009)]{Gopalswamy 2009}
Gopalswamy, N., M{\"a}kel{\"a}, P., Xie, H., Akiyama, S., \& Yashiro, S. 2009, \jgr, 114, A00A22

\bibitem[Gopalswamy et al. (2011)]{Gopalswamy 2011}
Gopalswamy, N., Nitta, N., Yashiro, S., et al. 2011, LWS/SDO-3/SOHO-26/GONG-2011 Workshop: Solar Dynamics and Magnetism from the Interior to the Atmosphere held October 31-November 4, 2011 in Stanford, CA. Online at http://sdo3.lws-sdo-workshops.org, id.22
(ADS link: https://ui.adsabs.harvard.edu/abs/2011sdmi.confE..22G/abstract )

\bibitem[Gopalswamy et al. (2013)]{Gopalswamy 2013b}
Gopalswamy, N., Xie, H., Akiyama, S., et al. 2013b, \apj, 765, L30

\bibitem[Gopalswamy et al. (2017)]{Gopalswamy 2017}
Gopalswamy, N., Yashiro, S., Akiyama, S., \& Xie, H. 2017, \solphys, 292, 65

\bibitem[Gopalswamy et al. (2015)]{Gopalswamy 2015}
Gopalswamy, N., Yashiro, S., Xie, H., Akiyama, S., \& M{\"a}kel{\"a}, P. 2015, \jgr, 120, 9221


\bibitem[Gopalswamy et al. (2001)]{Gopalswamy 2001}
Gopalswamy, N., Yashiro, S., Kaiser, M. L., Howard, R. A., \& Bougeret, J. L. 2001, \apj, 548, L91

\bibitem[Gopalswamy et al. (2002)]{Gopalswamy 2002}
Gopalswamy, N., Yashiro, S., Kaiser, M. L., Howard, R. A., \& Bougeret, J. L. 2002, \grl, 29, 106






\bibitem[Gosling (1991)]{Gosling 1991}
Gosling, J. T., McComas, D. J., Phillips, J. L., \& Bame, S. J.
1991, \jgr, 96, 7831












\bibitem[Huttunen et al. (2005)]{Huttunen 2005 ICME differnt identification}
Huttunen, K. E. J., Schwenn, R., Bothmer, V., \& Koskinen, H. E. J. 2005, AnGeo, 23, 625







\bibitem[Jian et al. (2018)]{JianLan 2018 ICME catalog}
Jian, L. K., Russell, C. T., Luhmann, J. G., \& Galvin, A. B. 2018, \apj, 855, 114








\bibitem[Kay \& Opher (2015)]{Kay 2015}
Kay, C., \& Opher, M. 2015, \apj, 811, L36


\bibitem[Kilpua et al. (2013)]{Kilpua 2013}
Kilpua, E. K. J., Isavnin, A., Vourlidas, A., Koskinen, H. E. J., \& Rodriguez, L. 2013, Ann. Geophys., 31, 1251







\bibitem[Kown et al. (2018)]{Kwon 2018 shock ratio}
Kwon, R. Y., \& Vourlidas, A. 2018, J. Space Weather Space Clim., 8, A08

\bibitem[Kwon et al. (2014)]{Kwon 2014 shock bubble}
Kwon, R. Y., Zhang, J., \& Olmedo, O. 2014, \apj, 794, 148



\bibitem[Liu et al. (2019)]{Liu Ying CME shock shape}
Liu, Y. D., Zhu, B., \& Zhao, X. H. 2019, \apjl, 871, 8L


















\bibitem[Longcope et al. (2007)]{Longcope 2007}
Longcope, D., Beveridge, C., Qiu, J., et al. 2007, \solphys, 244, 45







\bibitem[Makela et al. (2013)]{Makela et al. 2013}
M{\"a}kel{\"a}, P., Gopalswamy, N., Xie, H., et al. 2013, \solphys, 284, 59



\bibitem[Marubashi et al. (2015)]{Marubashi 2015}
Marubashi, K., Akiyama, S., Yashiro, S., et al. 2015, \solphys, 290, 1371



\bibitem[Miki\'{c} \& Linker (1994)]{Mikic 1994}
Miki\'{c}, Z., \& Linker, J. A. 1994, \apj, 430, 898








\bibitem[Odstrcil \& Pizzo (1999)]{Odstrcil 1999}
Odstrcil, D., \& Pizzo, V. J. 1999, \jgr, 104, 28255












\bibitem[Patsourakos et al. (2013)]{Patsourakos 2013}
Patsourakos, S., Vourlidas, A., \& Stenborg, G. 2013, \apj, 764,
125



\bibitem[Qiu et al.(2007)]{2007ApJ...659..758Q}
Qiu, J., Hu, Q., Howard, T.~A., \& Yurchyshyn, V.~B.\ 2007, \apj,
659, 758






\bibitem[Richardson \& Cane (2010)]{Richardson 2010 ICME catalog}
Richardson, I. G., \& Cane, H. V. 2010, \solphys, 264, 189










\bibitem[Riley et al. (1997)]{Riley 1997}
Riley, P., Gosling, J. T., \& Pizzo, V. J. 1997, \jgr, 102, 14677

\bibitem[Riley et al. (2006)]{Riley et al. 2006 Why ICME not MC}
Riley, P., Schatzman, C., Cane, H. V., Richardson, I. G., \& Gopalswamy, N. 2006, \apj, 647, 648











\bibitem[Song et al.(2014a)]{2014a ApJ...792L..40S Flux rope formation}
Song, H.~Q., Zhang, J., Chen, Y., \& Cheng, X.\ 2014, \apjl, 792,
L40



\bibitem[Song et al. (2019)]{3P structure in the EUV}
Song, H. Q., Zhang, J., Li, L. P., et al. 2019, \apj, 887, 124










\bibitem[Song \& Yao (2020)]{ICME composition review}
Song, H. Q., \& Yao, S. 2020, Sci. China Tech. Sci., 63, \\
http://engine.scichina.com/doi/10.1007/s11431-020-1680-y















\bibitem[Veronig et al. (2010)]{Veronig 2010}
Veronig, A. M., Muhr, N., Kienreich, I. W., Temmer, M., \& Vr\v snak, B. 2010, \apj, 716, L57



























\bibitem[Wu et al. (2011)]{WuCC ICME criteria}
Wu, C. C., \& Lepping, R. P. 2011, \solphys, 269, 141

\bibitem[Xie et al. (2013)]{Xie et al. 2013}
Xie, H., Gopalswamy, N., \& St. Cyr, O. C., 2013, \solphys, 284, 47





\bibitem[Yashiro et al. (2013)]{Yashiro 2013}
Yashiro, S., Gopalswamy, N., M{\"a}kel{\"a}, \& P., Akiyama, S. 2013, \solphys, 284, 5









\bibitem[Zhang et al. (2013)]{Zhang 2013}
Zhang, J., Hess, P., \& Poomvises, W. 2013, \solphys, 284, 89

\bibitem[Zhang et al.(2007)]{2007JGRA..11210102Z}
Zhang, J., Richardson, I.~G., Webb, D.~F., et al.\ 2007, \jgr,
112, A10102




\bibitem[Zurbuchen \& Richardson (2006)]{ICME criteria}
Zurbuchen, T. H., \& Richardson, I. G. 2006, \ssr, 123, 31



\end{thebibliography}
\end{document}